\documentclass[twocolumn]{aastex631}

\usepackage{url}
\usepackage{amsmath}
\usepackage{amssymb}
\usepackage{graphicx}
\usepackage[version=4]{mhchem}
\usepackage{soul}

\newcommand{\gcc}{g\,cm$^{-3}$}
\newcommand{\kgcc}{$kg/m^3$}

\shorttitle{Serpentinization in dwarf planets}
\shortauthors{Farkas-Takács \& Kiss}

\graphicspath{{./}{figures/}}

\begin{document}

\title{The impact of serpentinization on the initial conditions of satellite forming collisions of large Kuiper belt objects}

\author[0000-0001-5531-1381]{Anik\'o Farkas-Tak\'acs}
\affiliation{Konkoly Observatory, Research Centre for Astronomy and Earth Sciences, HUN-REN, Konkoly Thege 15-17, H-1121~Budapest, Hungary}
\affiliation{CSFK, MTA Centre of Excellence, Budapest, Konkoly Thege Miklós út 15-17, H-1121, Hungary}

\author[0000-0002-8722-6875]{Csaba~Kiss}
\affiliation{Konkoly Observatory, Research Centre for Astronomy and Earth Sciences, HUN-REN, Konkoly Thege 15-17, H-1121~Budapest, Hungary}
\affiliation{CSFK, MTA Centre of Excellence, Budapest, Konkoly Thege Miklós út 15-17, H-1121, Hungary}
\affiliation{ELTE E\"otv\"os Lor\'and University, Institute of Physics and Astronomy, P\'azm\'any P. st. 1/A, 1171 Budapest, Hungary}

\begin{abstract}

Kuiper belt objects are thought to be formed at least a few million years after the formation of calcium–aluminium-rich inclusions, at a time when the $^{26}$Al isotope -- the major source of radiogenic heat in the early Solar System -- had significantly depleted. The internal structure of these objects is highly dependent on any additional source that can produce extra heat in addition to that produced by the remaining, long-lasting radioactive isotopes. In this paper, we explore how serpentinization, the hydration of silicate minerals, can contribute to the heat budget and to what extent it can modify the internal structure of large Kuiper belt objects. We find that the extent of restructuring depends very strongly on the start time of the formation process, the size of the object, and the starting ice-to-rock ratio. Serpentinization is able to restructure most of the interior of all objects in the whole size range (400-1200\,km) and ice-to-rock ratio range investigated if the process starts early, $\sim$3\,Myr after CAI formation, potentially leading to a predominantly serpentine core much earlier than previously thought ($\leq$5\,Myr vs. several tens of million years). While the ratio of serpentinized material gradually decreases with the increasing formation time, the increasing ice-to-rock ratio, and the increasing start time of planetesimal formation in the outer solar system, in the case of the largest objects a significant part of the interior will be serpentinized even if the formation starts relatively late, $\sim$5\,Myr after CAI formation. Therefore it is feasible that the interior of planetesimals may have contained a significant amount of serpentine, and in some cases, it could have been a dominant constituent, at the time of satellite-forming impacts. 


\end{abstract}
\keywords{Dwarf planets (419) --- Solar system (1528)}



\section{Introduction}

Serpentinization of ultramafic rocks is a water-dependent geochemical process, which includes all transformations involving hydrolyzation of olivine \ce{((Mg,Fe)2SiO4)} and pyroxenes \ce{((Mg,Fe)SiO3)}. 
During the reaction, a variety of gas and fluid species and rocks containing serpentine-dominated minerals are formed. 
According to some meteorite samples, inside the small bodies of the solar system, there may have been chemical processes. These processes were able to significantly alter the lithological and mineralogical characteristics and as an additional heat source, they contributed to the internal heat production. Such transformations are explained by serpentinization reactions which may have had an important role in sculpting the internal structure of large planetesimals.

%


Serpentine group materials seem to be common in the Solar system. The global reflectance spectrum of Ceres shows a serpentine abundance of 10-25\% \citep{Prettyman2017}, the 'dark material' on the surface of Vesta has also been identified as serpentine, likely deposited from CM chondrite-type impactors \citep{Nathues2014}, and the serpentine is the primordial candidate to explain the reflectance spectra of trans-Neptunian objects \citep{Protopapa2009}.

If the initial conditions allow the serpentinization reaction, and differentiation can take place, it results in a different internal structure of larger bodies \citep{Cioria2022}, e.g. in a larger serpentine core with $\rho\,\sim$\,2.4\,\gcc\, rather than a smaller rocky core with $\rho\,\sim$\,3.2\,-\,3.5\,\gcc. 
The serpentinization process has been invoked to explain the interior of Pluto \citep[see e.g.][]{Vance2007}, and it has been proposed by \citet{Dunham2019} that Haumea's core underwent serpentinization in the past. 
\citet{Noviello2022} presented a detailed internal evolution model of Haumea including the serpentinization process that led to core growth and increased Haumea's moment of inertia on a few hundred million-year time scale, assuming that the core forming event occurred $\sim$50\,Myr after CAI formation.

The efficiency of the serpentinization process is very sensitive to the actual local temperature which itself strongly depends on the amount of heat-producing radiogenic nuclei ($^{26}$Al) in the early stages of the evolution.
As it was shown by \citet{Monnereau2023} for the inner Solar system, if the accretion time is less than 1.5\,Myr after the Ca–Al-rich inclusion (CAI) formation then the whole body can be differentiated, but if the accretion is completed later, it can differentiate only partially or not at all. 




Almost all dwarf planets in the Kuiper belt have at least one known satellite \citep{Noll2020,Holler2021}. 
These systems are very different from the smaller binaries where the components are often found as nearly equal-sized with a large relative separation, suggesting different formation conditions for the two types of objects.
While different formation scenarios are possible \citep[e.g. mass ejection due to fast rotation, see][]{Noviello2022}, the satellites of the outer solar system dwarf planets were most likely created by energetic collisions between massive planetesimals \citep{Holler2021}. 



In the currently accepted scenario large (100-1000\,km) Kuiper belt objects formed near 25\,AU via streaming instability by a few million years after CAI formation. The initial small dust balls continued growth via pairwise collisions and pebble accretion can produce a size distribution of large bodies in $\sim$3\,Myr that consistently explains the current properties of the Kuiper belt \citep{Johansen2015,Morbidelli2020}.
Collisions between large bodies can continue for a few tens of million years. 
For example, a timeline put forward for the Pluto-Charon system \citep{Canup2021} suggests that the accretion process in the outer solar system started in 3-7\,Myr after CAI formation, lasted for 1-2\,Myr, and the binary formation occurred in 7-30\,Myr after CAI formation. 

A key parameter that strongly affects impact outcome is the composition of the progenitors and the distribution of ice and rock in the interiors at the time of the collisions \citep{Arakawa2019,Canup2021}. \citet{Canup2021} 
argue that short-lived radiogenic nuclei do not play a major role in internal heating -- necessary for differentiation to initiate -- based on a likely timescale of several million years of progenitor accretion \citep{Morbidelli2020}, and a lack of widespread observed thermal processing of outer solar system materials \citep{Neveu2019}. However, heat from the decay of longer-living radiogenic nuclei can initiate internal ice-rock differentiation within tens of millions of years \citep{Desch2017,Canup2021}, especially for large (Pluto-sized) objects. If this is the case, differentiated progenitors should have existed only toward the end of the binary formation window ($\sim$20-30\,Myr). 

In giant impact simulations, it is often assumed that one of the main constituents of the progenitor material is serpentine, either in an undifferentiated, uniform composition or mixed with ice. Impacts that produce large intact moons were first seen in Pluto-Charon simulations using uniform serpentine compositions \citep{Canup2005}. \citet{Arakawa2019} found that most secondaries around the presently known large TNOs are likely intact moons and not disk-origin satellites, putting constraints on the initial composition and structure of the progenitor and the impactor.
The formation of serpentine may also serve as a significant heat source \citep{Farkas2022} that may also initiate differentiation even when $^{26}$Al ran out, in much shorter timescales than the longer-lasting radionuclides. 
As the outcome of these giant collision simulations critically depends on the initial composition and structure, it is important to investigate under which conditions serpentinization could reform the interior of the progenitors and produce heat sufficient for differentiation within the limited time window of binary formation after CAI formation.

In this paper, we investigate this question by performing coupled internal heat, accretion, and chemical evolution (serpentinization) simulations of planetesimals in the early solar system. We study the first few million years of evolution assuming different evolutionary conditions of the planetary embryos and determine how the composition, internal structure, and temperature distribution changes with time in this early era, considering serpentinization. 
Our results also provide consistent initial conditions for giant collisions, depending on the formation conditions and history of the progenitors (size, original composition, and formation time).

\section{Methods}


To model the thermal evolution of planetesimals, we used the model presented in \citep{Farkas2022} that is able to follow the thermal evolution of these objects considering heat from radiogenic decay and accretion, and follows the chemical and thermodynamical changes due to the serpentinization process. This model takes into account the presence of interfacial liquid water below the melting point of ice \citep{Gobi2017}, thus the reaction can also take place at lower temperatures. In our simulations, we considered objects that correspond to the largest bodies (D\,$\approx$\,800-2400\,km) in the Kuiper belt today. The objects are assumed to be spherical a valid assumption for slowly rotating bodies in the Kuiper belt with sizes D\,$\gtrsim$\,400\,km  \citep{PotatoRadius}, and the simulation starts with a small object that is 1/20 of the final size, \textit{R$_f$\,=\,[400, 600, 800, 1000, 1200]\,km}. The mass is gradually increasing with equally distributed 1/20-mass steps, assuming that accretion is spherically symmetric (either from pebble accretion or from smaller collisions) and that the same material (i.e. same ice-to-rock ratio, see below) is accreted in a 1\,Myr time. 

The start time of the simulation, \textit{t$_s$\,=\,[3, 4, 5]\,Myr} after the time of CAI formation, sets the amount of $^{26}$Al available to produce radiogenic heat in the rocky component, also considering the decreasing amount of $^{26}$Al in the subsequent accretion steps. The whole simulation took 1.5 million years and during the last simulated half million years, heat is provided by serpentinization and the decay of radioactive nuclei. The model uses a pressure-dependent porosity \citep{Yasui2009}, and the distribution of the components is homogenous and isotropic at the start of the process. In the simulations, the material of the planetesimal has seven basic components: silicate rocks such as forsterite, enstatite, serpentinite, and other non-reactive rocks; \ce{H2O} which is present in three phases: liquid, solid, and gas. 
Based on the meteorite samples, we used a non-reactive rock content of 14\%  in the planetesimals with a density of 3630\,\kgcc\, as in earlier studies \citep{Farkas2022, Cohen2000}.
The heat released during accretion provides the initial temperature and after accretion, the main internal heat source is the decay of radionuclides, in addition to the heat potentially obtained from serpentinization \citep[see][for a more detailed explanation]{Farkas2022}. 

One critical parameter in these simulations is the density of the material accreted. We assume that this material is made of an {\it icy} and a {\it rocky} component, and the bulk density is a consequence of their mixing ratio. This also determines the amount of radionuclide present, as these can be found in the rocky component only \citep[for the actual concentration considered see][]{Farkas2022}. 

\citet{Barr2016} assumed a uniform primordial density of $\sim$1.8\,\gcc, which may represent the bulk composition of the primordial material from which the KBOs accreted. Pluto has a similar density as well.
However, some of the largest dwarf planets have densities notably higher than this $\gtrsim$\,2\,\gcc, including Eris, Makemake, Haumea, and Triton \citep{Holler2021Eris,Parker2018,Noviello2022,Dunham2019}.
To explain the high density of these objects the only feasible scenario is a high primordial density. Collisions cannot remove enough material \citep{Brown2012} to increase the densities to these high values, and evaporation of volatiles would require a significant amount of extra heat e.g. from tidal interaction which may only be conceivable in the case of Triton \citep{Barr2016}. 

Also, $\sim$1000\,km-size Kuiper belt objects have densities of $\sim$1.6\,\gcc \citep[e.g. Orcus and Quaoar,][]{Brown2018,Morgado2023}, below the primordial density suggested by \citet{Barr2016}. The porosity should already have a small effect in this size range \citep{Brown2013a,Fernandez2020}, indicating that there may be primordial density differences between the dwarf planets. To account for this possible difference, we considered three types of primordial densities: Eris-like ($\rho$\,=\,2.4\,\gcc) that corresponds to an initial olivine-to-water mixing ratio of $\nu_0$\,=\,0.49, Pluto-like ($\rho$\,=\,1.8\,\gcc, $\nu_0$\,=\,0.11) and Orcus-like ($\rho$\,=\,1.53\,\gcc, $\nu_0$\,=\,0.04). A similar range of primordial densities (rock mass fraction of 0.2--0.8) was used \citep{Loveless2022} to study the long-term internal structure and temperature evolution of icy bodies. 

Our simulations were performed for all combinations of primordial density $\rho_p$\,=\,[1.53, 1.8, 2.4]\,\gcc, formation start time 
t$_s$\,=\,[3, 4, 5]\,Myr, and final effective radius R$_f$\,=\,[400, 600, 800, 1000, 1200]\,km. The results are presented below. 

\section{Results}


\subsection{Thermal evolution}

\begin{figure}[!ht]
\begin{center}
\includegraphics[width=\columnwidth]{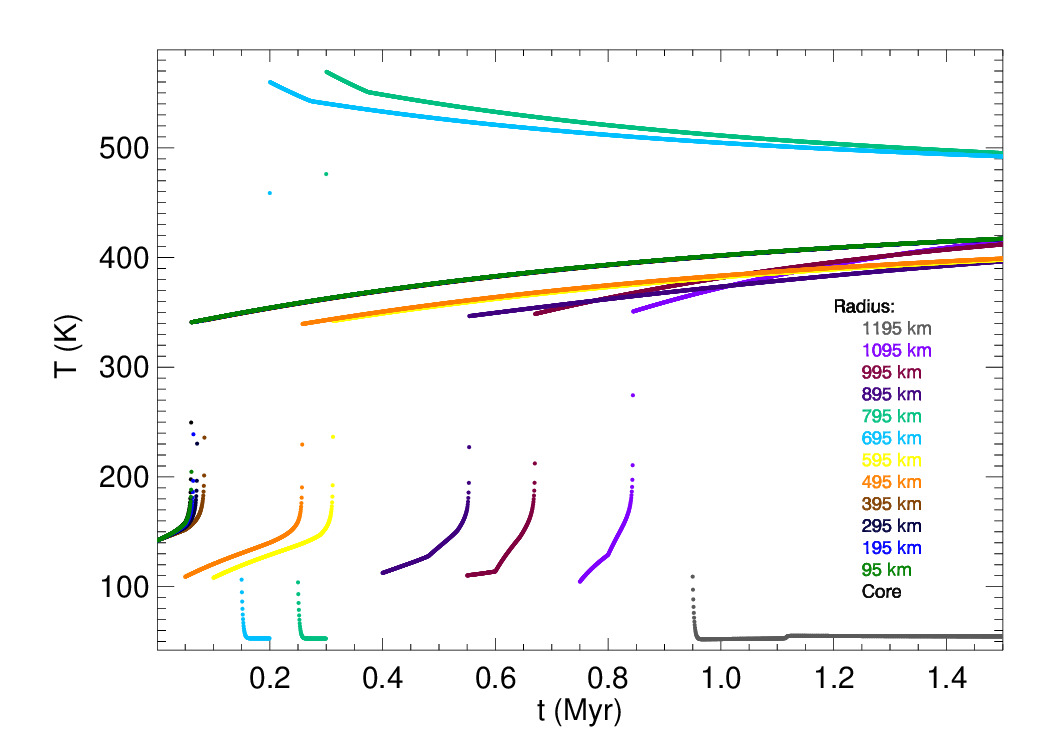}
\includegraphics[width=\columnwidth]{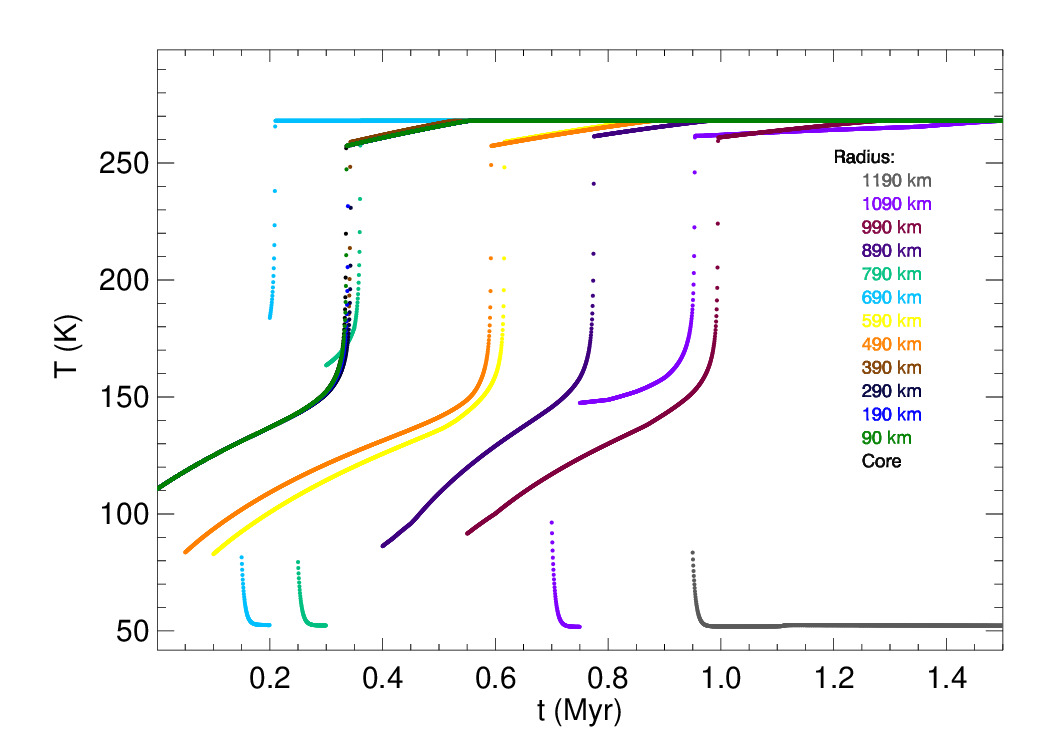}
\includegraphics[width=\columnwidth]{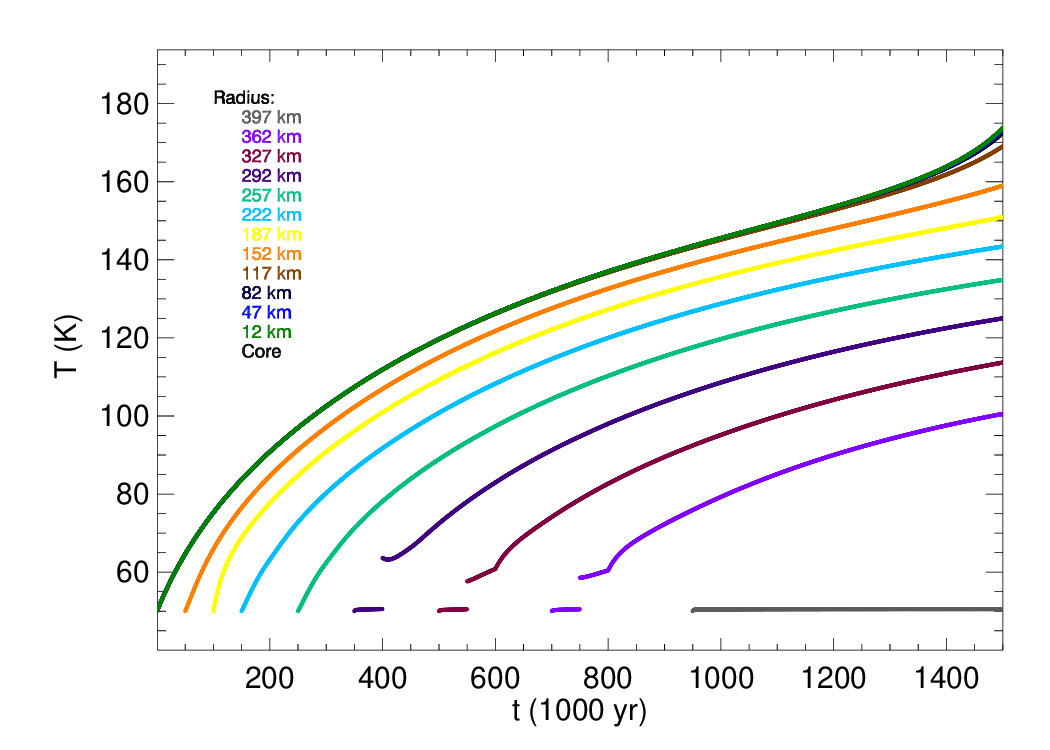}
\caption{Illustration of the thermal evolution of the inner layers of planetesimals. Each curve corresponds to a layer of a specific depth, as indicated by the inserts.  
Upper: [$\rho_s$\,=\,2.4\,\gcc, R$_s$\,=\,1200\,km, t$_s$\,=\,4\,Myr] ($\sim$Eris); 
middle: [$\rho_s$\,=\,1.8\,\gcc, R$_s$\,=\,1200\,km, t$_s$\,=\,4\,Myr] ($\sim$Pluto); 
bottom: [$\rho_s$\,=\,1.53\,\gcc, R$_s$\,=\,400\,km, t$_s$\,=\,4\,Myr] ($\sim$Orcus).
}
\label{fig:heatevol}
\end{center}
\end{figure}

As a result of the continuous accretion (lasting 1 million years), the temperature distribution will be inhomogeneous at the start of the serpentinization reaction.
Figure \,\ref{fig:heatevol} illustrates the thermal evolution of the inner layers for three different cases. In these examples, the formation started 4\,Myr after CAI, and we used the size of the namesake dwarf planet of a specific composition, i.e. $R_s$\,=\,1200\,km and $\rho_p$\,=\,2.4\,\gcc\, ('Eris'), 
$R_s$\,=\,1200\,km and $\rho_p$\,=\,1.8\,\gcc\, ('Pluto') and 
$R_s$\,=\,400\,km and $\rho_p$\,=\,1.53\,\gcc\, ('Orcus'). 
Those curves in the figures that first show quick cooling and then sudden warming are initially the surfaces of the planetesimal at a certain moment of the evolution.
 As accretion progresses, these cool surfaces become an inner layer and are heated by energy released during accretion.

\begin{figure}
\begin{center}
\resizebox{8.5cm}{!}{\rotatebox{0}{\includegraphics{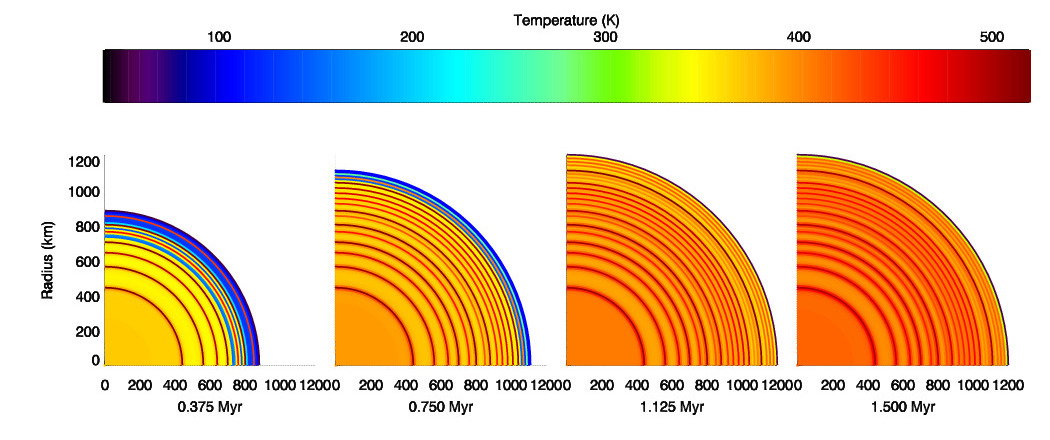}}}
\resizebox{8.5cm}{!}{\rotatebox{0}{\includegraphics{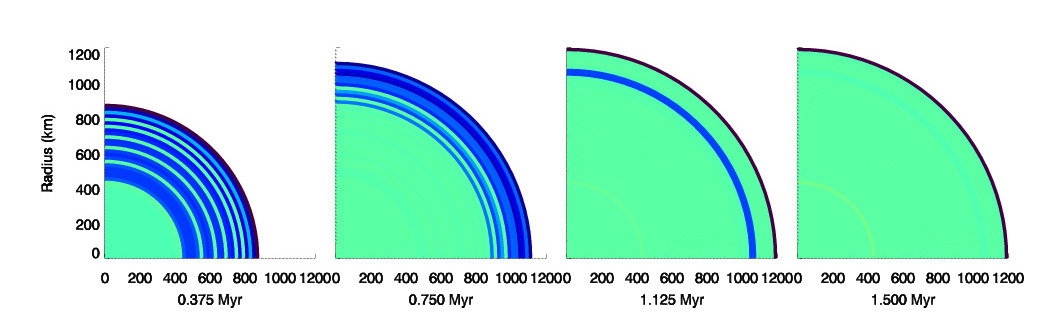}}}
\resizebox{8.5cm}{!}{\rotatebox{0}{\includegraphics{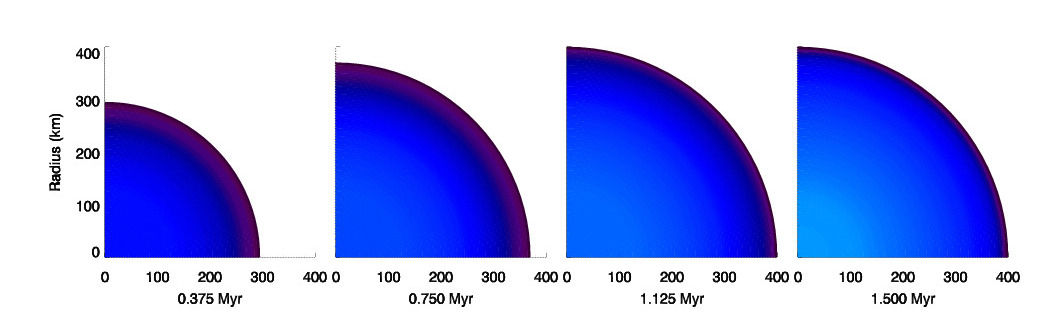}}}
\caption{Internal temperature distribution of 'Eris' (upper), 'Pluto' (middle), and 'Orcus' (bottom) with their size at different times during the formation and in half a million years after the stop of the accretion, when the formation starts 4 million years after CAI formation. The large green area visible on the 'Pluto' example is the temperature of the melting point of \ce{H2O} ice, where the heat produced is used to melt the crust over a long period of time.}
\label{fig:heatevol_full}
\end{center}
\end{figure}

In the case of bodies with Orcus-like densities with a high \ce{H2O} content only the temperature increase resulting from radioactive decay is visible (bottom panel in Fig.~\ref{fig:heatevol}). In contrast in the two other cases, serpentinization provides a significant amount of heat to the entire system.
This is shown by the sudden warming in the simulations of 'Pluto' and 'Eris' at the time of the effective start of the reaction.
The reason for this behavior is the dependence of pressure on the object size that strongly determines the reaction rate, and is obviously smaller for a smaller object. Another key factor in the reaction rate is the temperature which simply does not reach the value necessary for the efficient start of the reaction in the case of smaller and icier objects, like Orcus. While the temperature in 'Pluto' reaches the melting point of water ice during the studied period, and some layers of 'Eris' heat up to 500\,K, the temperatures of all layers of 'Orcus' remain below the melting point. The three examples clearly show that a larger temperature increase can be achieved with a higher rock content (fig.~\ref{fig:heatevol}, \ref{fig:heatevol_full}).

An interesting feature is the layered structure of 'Eris' and 'Pluto' observed when we examined the evolution of the temperature distribution (see Fig.~\ref{fig:heatevol_full}). As the planetesimals grow by accretion, warmer layers are formed, and in these layers, the chemical process 
can begin and progress faster, providing an extra heat source. These layers will be significantly warmer than their surroundings, 
and thus they transfer heat to the colder layers, where the reaction can also start, causing a local increase in temperature, 
until finally, almost the whole object can be serpentinized.
As \citet{Farkas2022} pointed out that, if the serpentinization reaction can start  effectively it can be completed in a few ten thousand years even in the case of a late formation, because the reaction itself produces enough heat to continue. Thus, in those simulations where the process has started, but serpentinization has not been completed, it will most likely complete in several tens of thousands of years.
Figs~\ref{fig:heatevol} and \ref{fig:heatevol_full} clearly show that in the case of 'Orcus', 
serpentinization does not play a significant role in the heat evolution, since it cannot start effectively, and therefore the layered structure is also not visible in the internal temperature distribution.

\subsection{Lithological and mineralogical evolution}
 
\begin{figure}[ht!]
\begin{center}
\includegraphics[width=\columnwidth]{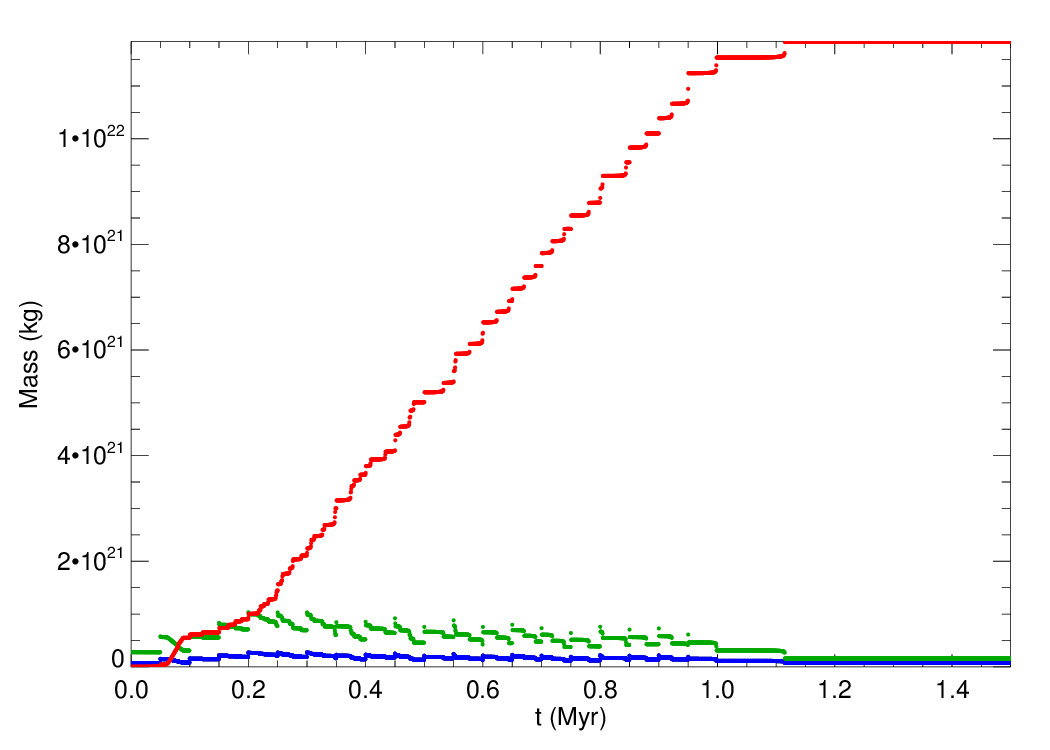}
\includegraphics[width=\columnwidth]{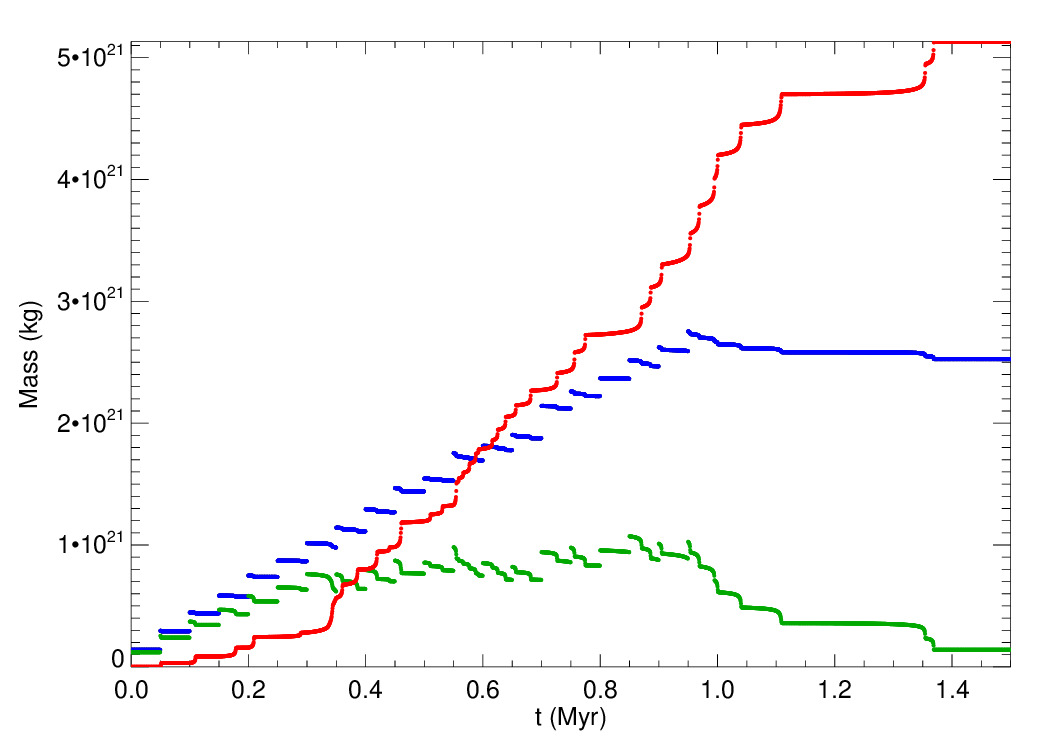}
\includegraphics[width=\columnwidth]{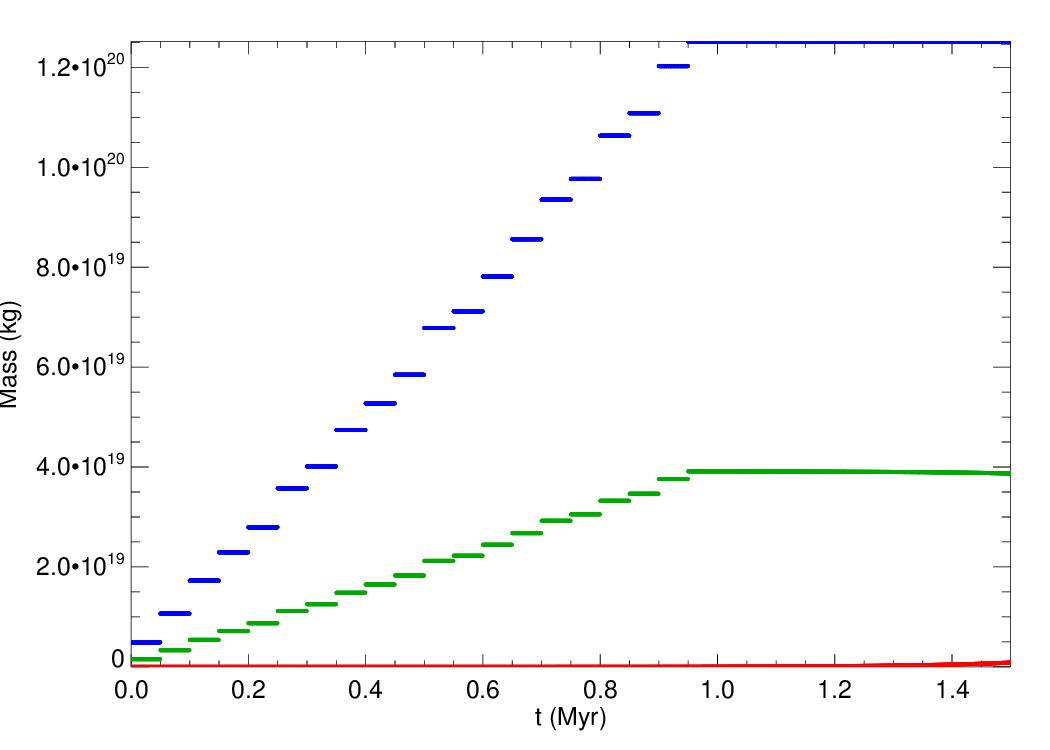}
\caption{Evolution of \ce{H2O} (blue; including solid, liquid, and vapor phase), olivine (green), and serpentine (red) mass of the test object as a function of time, for the three simulations cases also presented in Fig.~\ref{fig:heatevol}. 
Upper panel: [$\rho_s$\,=\,2.4\,\gcc, R$_s$\,=\,1200\,km, t$_s$\,=\,4\,Myr] ($\sim$Eris); 
middle: [$\rho_s$\,=\,1.8\,\gcc, R$_s$\,=\,1200\,km, t$_s$\,=\,4\,Myr] ($\sim$Pluto); 
bottom: [$\rho_s$\,=\,1.53\,\gcc, R$_s$\,=\,400\,km, t$_s$\,=\,4\,Myr] ($\sim$Orcus).}
\label{fig:massevol}
\end{center}
\end{figure}

In addition to its significant contribution to internal heat production, serpentinization has a very important role in internal lithological and mineralogical changes. In the case of Pluto, it was previously determined that its core was probably hydrated during its early evolution \citep{Denton2021}.
Depending on the composition, even the total mass of a planetesimal can be transformed by chemical processes. However, the surface is not able to serpentinize, because the heat will not be sufficient due to the rapid cooling. In the simulations for 'Eris' (whose composition may be close to the stoichiometric ratio of the considered serpentinization reaction equation), the initial temperature approached the temperature required for the effective initiation of the serpentinization reaction, and then quickly reached it with the help of radioactive decay in the early phase of the accretion. The amount of serpentine shows a continuous and almost uniform increase (fig.\,\ref{fig:massevol} red line on the top panel) until the olivine runs out. When the conditions are right (temperature, pressure, etc.), the reaction can start even during the accretion.
Small jumps indicate the accretion phases when the reaction rate suddenly increases due to the amount of heat released. The amount of olivine and water initially increases for a short time period as a result of accretion, after which serpentinization can start efficiently and the reaction extends to the entire object. From here on, we can see a decrease despite the accretion, as the reaction consumes reactants faster than the rate at which accretion produces them.

In the case of 'Orcus', only the continuous growth of the amount of components can be observed due to the accretion. The mass of serpentine starts to increase at the end of the examined period (Fig.~\ref{fig:massevol}, bottom panel), but due to the low temperatures, the small amount of radiogenic heat, and the stop of accretion the serpentinization efficiency will remain low, leaving the final bulk composition close to the initial one. 
'Pluto' (Fig.~\ref{fig:massevol}, middle panel) seems to follow a different evolution. The accretion phases are clearly defined here as well, as in the case of 'Eris', by the sudden increase in the mass of the serpentine. However, between these stages, there are periods when there is no significant reaction rate. The reason for this is that in the case of 'Pluto', the initial parameters created a state in which the system is very close to effectively starting the reaction, and any additional heat (e.g. that from accretion) case easily pushes a layer into a fast reaction stage. These objects, even with the primordial density smaller than in the case of 'Eris', can reach a state when the maximum amount of serpentine is produced, considering the initial composition. 

In studies of the evolution of Saturn's moons \citep{Neveu2017, Neveu2019} or Charon \citep{Desch2017} serpentinization appears as hydration of the rock and as a source of heat. However, these studies do not take into account the effect of interfacial water below the melting point of ice, which would place the start of the reaction earlier, even in the formation phase, thus contributing to the differentiation process. As a consequence of this, today's differentiated structure can even be described as a later origin.

\subsection{Bulk composition at the end of the formation phase}

One of the key results that can be obtained from our simulations is the bulk composition of the planetesimals at the end of the present simulations, i.e. in 1.5\,Myr after the start.
We describe this simply by comparing the 'rock', 'water', and 'serpentine' content of the planetesimals. Here 'water' refers to water in any phase (solid, liquid, or vapor), and 'rock' refers to any material that is not 'water' and not transformed to hydrated minerals (serpentine) by the end of the simulations. 
We emphasize that at this stage, even if the conditions are feasible, convection may not have separated the components, but it could eventually lead to a fully differentiated body in a later stage. 

\begin{figure}[!ht]
\begin{center}
\includegraphics[width=8.5cm]{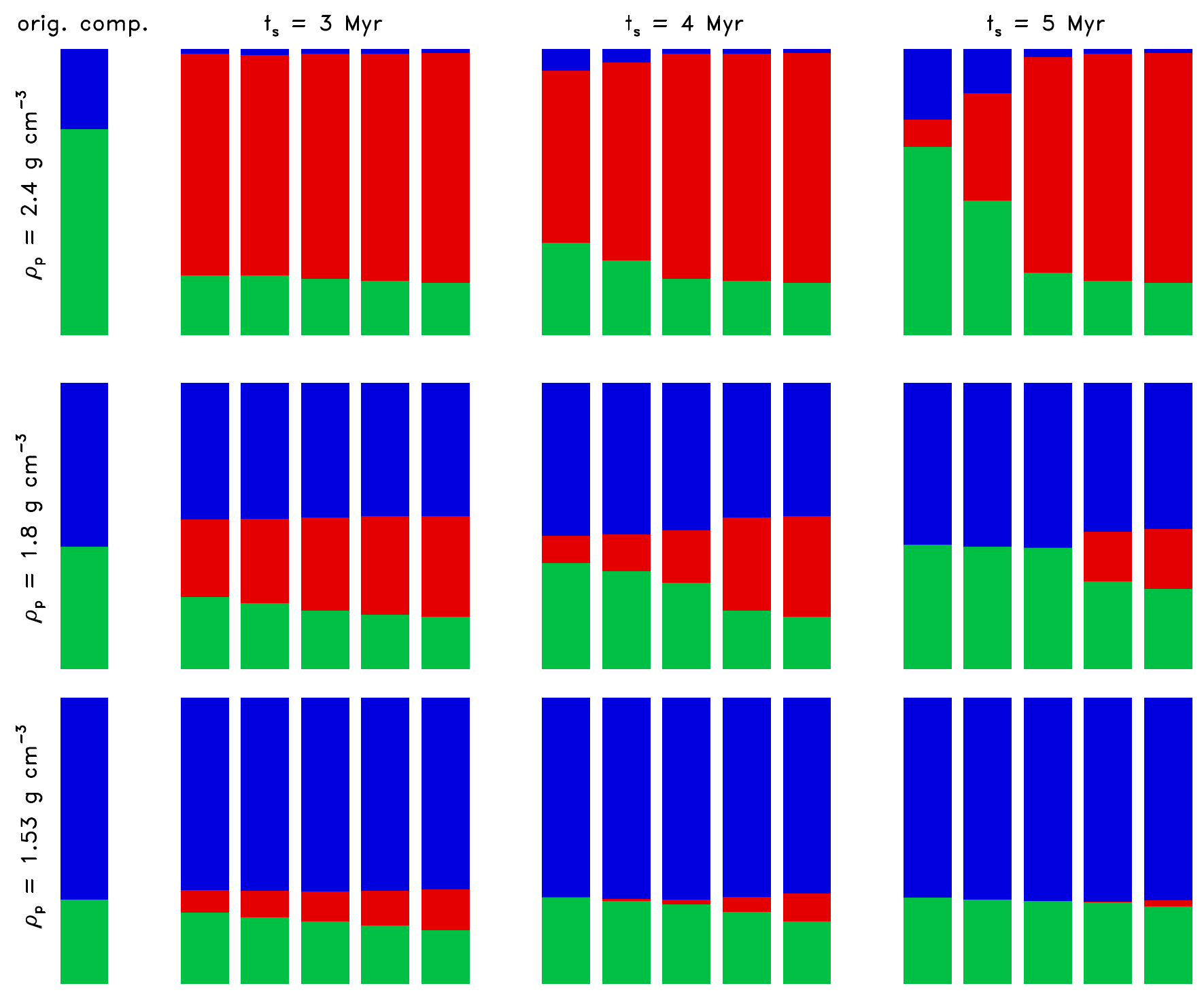}
\caption{Bulk composition of the planetesimals at the end of the simulations, 0.5\,Myr after the end of the accretion process. Blue, green, and red colors correspond to water ice, olivine and/or non-reactive rock, and serpentine, respectively. The first bar in each row represents the initial composition (water/olivine ratio). The bars correspond to the initial density and formation time as indicated, and all five-bar blocks correspond to R$_s$\,=\,[400, 600, 800, 1000, 1200]\,km, from left to right. 
}
\label{fig:serp3}
\end{center}
\end{figure}

As seen in Fig.~\ref{fig:serp3}, for low primordial densities ($\rho_p$\,$\approx$\,1.5\,\gcc) the effect of serpentinization is small, and only a small amount of serpentine forms, even for the earliest starting time and the largest objects. For the latest starting time almost no serpentine can form, and the final bulk composition remains essentially the same as the starting one, 
and these objects may remain undifferentiated. 
For high primordial densities ($\rho_p$\,=\,2.4\,\gcc) and at early and intermediate staring times -- $t_s$\,=\,3 and 4\,Myr -- most of the potentially available material is consumed by serpentinization, leaving only a small amount of 'free' water in the system. In the $t_s$\,=\,5\,Myr case the process is strongly size-dependent: large objects transform most of their material to serpentine, while the process is far less effective in the case of the smallest planetesimals considered in our simulations. Note that we assumed a certain amount of non-reacting material in the simulations, i.e. there is always a small amount of 'rocky' component that remains in the system, even if the efficiency of the serpentinization is 100\%. 

For intermediate primordial densities ($\rho_p$\,=\,1.8\gcc) the efficiency depends strongly both on $t_s$ and $R_f$. Large objects ($R_f$\,$\geq$\,1000\,km) form some amount of serpentine even if the formation starts late, while the smallest may not form any or a notably smaller amount than the large ones at the same starting time. 

\begin{figure}[!ht]
\begin{center}
\includegraphics[width=\columnwidth]{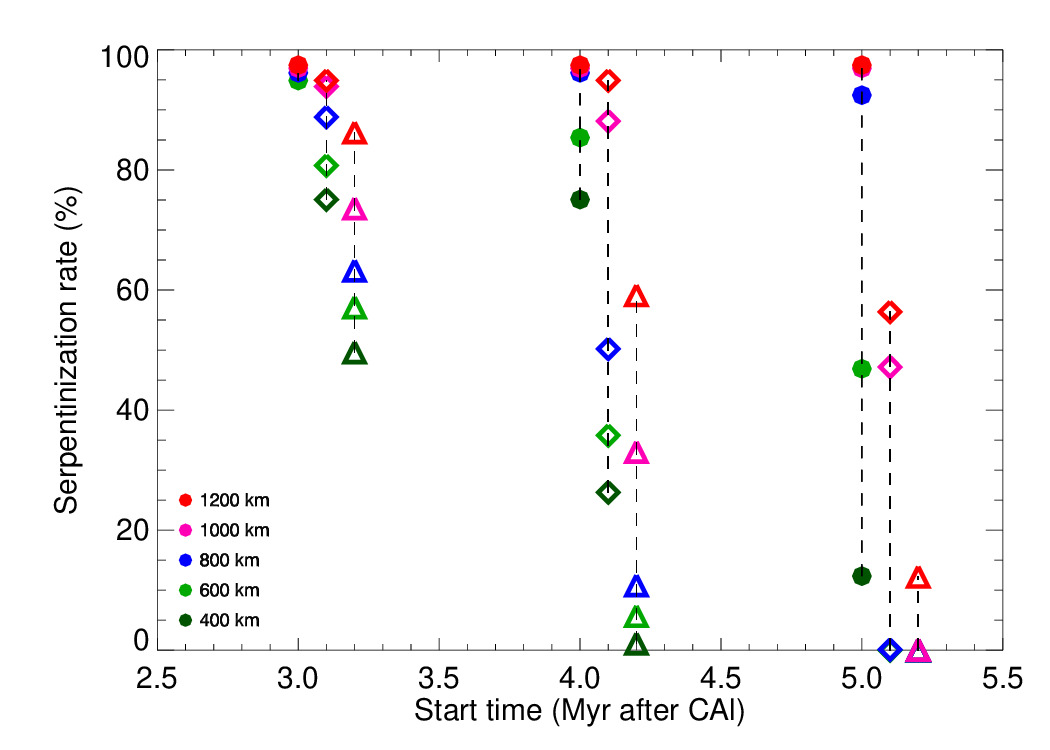}
\caption{Serpentinization rates for different sizes in Eris- (full circle), Pluto- (diamond), and Orcus-type (triangle) objects, assuming a 1 Myr accretion starting 3-5 Myr after CAI formation. For clarity, Pluto-type and Orcus-type runs are shifted by 0.1 and 0.2\,Myr in the figure.}
\label{fig:serp_comp}
\end{center}
\end{figure}

Some more insights can be gained on the efficiency of the process by looking at the final serpentinization rates at the end of our simulations, considering only the substances which could have participated in the reaction
(Fig.~\ref{fig:serp_comp}). These rates are very close to 100\% for high primordial densities and large sizes at any formation time, and they are in general high for early ($t_s$\,=\,3\,Myr) formation. 
While our simulations stop at 0.5\,Myr after the end of accretion, the high temperatures associated with early formation indicates even in the case of smaller and lower density objects that the serpentinization reaction will go on, and eventually lead to an even higher serpentinization rate. 
For later formation times and lower primordial densities, the efficiency is substantially reduced, and almost no serpentine can be formed (note that almost all points are $\sim$zero for $t_s$\,=\,5\,Myr and $\rho_p$\,=\,1.53\,\gcc). 

\section{Discussion and Conclusions}

In our simulations, we considered planetesimals with different sizes from a radius of 400 to 1200\,km which approximately covers the known dwarf planet size range, and also different accretion start times (3, 4, and 5 million years after CAI formation). The reaction rate strongly depends on the temperature and the lithospheric pressure, and therefore, as expected, on the size, the time of formation, and the initial composition. The interior composition of bodies with a high rock content was more likely to be partially or completely transformed even in the case of late formation since they had a higher initial temperature from accretion and higher heat production from radioactive decay. In objects with low-density 
, the time of formation has a much greater influence on the chemical reactions and the final composition. The correlation is also seen with size, because the larger an object, the easier it is for serpentinization to start. The final size of the planetesimal is an especially important factor for serpentinization efficiency in the case of later formation times. 

The formation of Sputnik Platina on Pluto was simulated using impact modeling by \citet{Denton2021}. Their results were strongly influenced by the thickness of the subsurface ocean and the composition of the core. The simulations that best reproduce the observed terrains suggested that Pluto may have had an ocean $\geq$150\,km thick at the time of the impact and a core that was hydrated i.e. composed mainly of serpentine. 
While previous calculations \citep[e.g.][]{Wakita2011} suggested that icy planetesimals, formed at 2.4\,Myr after CAI formation, may not reach the melting temperature of ice, regardless of their sizes and initial temperature, our simulations clearly show that these temperatures can indeed be reached for a notable range of initial conditions, and even in the case of a late formation, start at 5\,Myr after CAI formation. 

\citet{Loveless2022} performed a thorough investigation of the long-term (4.5\,Gyr) internal structure evolution of icy bodies, including serpentinization. In their calculations they did not consider the heat from short-lived radionuclides ($^{26}$Al) therefore their results are not directly comparable with our results on the early evolution of these planetesimals. In all their models, even with a high rock mass ratio and large size, temperatures remain low throughout the whole object in the first $\sim$100\,Myr, and the formation of the core is significantly delayed, mainly due to the lack of the heat from short-lived radionuclides. 
Eventually serpentine forms in the core in their models as well, but on a significantly longer timescale. It also means that the \citet{Loveless2022} models would prevent serpentine to be involved in satellite-forming giant collisions, even if they occur several tens of million years or even a hundred million years after CAI formation. 
Our simulations, however, show that serpentine can form quickly under a relatively wide range of conditions, at least for the largest objects.
It is therefore feasible that at the time of the satellite forming impacts planetesimals may already have contained a significant amount of serpentine, even if these events happened early, $\sim$5--7\,Myr after CAI formation. 

These results can be applied directly to the Eris-Dysnomia system. 
Eris's rotation has recently been found to be tidally locked, synchronized with the orbital period of its satellite, Dysnomia \citep{Szakats2023,Bernstein2023}. Tidal evolution calculations suggest that to reach this state, Dysnomia has to be large ($\sim$700\,km) and dense (1.8--2.4\,\gcc). 
As indicated by our simulations most of Eris' interior may have serpentinized quickly, even if the formation started relatively late. If during the satellite forming impact, the temperatures remain below the dehydration temperature of 873\,K \citep{Wakita2011} the satellite may form mainly of serpentine, setting the density to a value indicated by tidal constraints, even assuming some moderate level of porosity.
A recent study by \citet{BB23}, however, indicates that the density of Dysnomia could be smaller, $\lesssim$1.2\,\gcc, implying a Dysnomia–Eris mass ratio of q\,=\,0.0085. In giant impact models, this q value is in the transition region between those of intact impactor and reaccreted disk material outcomes for the satellite. Our results suggest that if Dysnomia formed from reaccreted disk material it should have formed very early after the end of Eris' accretion, before serpentinization could have taken place at the outer layers of Eris. In addition to simulations of internal evolution, size/mass determinations of large Kuiper belt object satellites will be the key to constrain and understand the early evolution of dwarf planets and the formation of their satellites in the trans-Neptunian region.



\section*{Acknowledgements}
This research has been supported by the K-138962 project of the National Research, Development and Innovation Office (NKFIH, Hungary). We are indebted to our referee for the thorough review and the useful comments which have improved the paper considerably.





\bibliography{ref}{}
\bibliographystyle{aasjournal}

\end{document}